# Topology and spin dynamics in magnetic molecules

S. Carretta[1], P. Santini[1], G. Amoretti[1], M. Affronte[2], A. Ghirri[2],
I. Sheikin[3], S. Piligkos[4], G. Timco[4] and R.E.P. Winpenny[4]

[1] I.N.F.M. and Università di Parma, Parco delle Scienze, 43100 Parma, Italy.
[2] I.N.F.M. - S[3] National Research Center on nanoStructures and bioSystems at Surfaces and Dipartimento di Fisica,
Università di Modena e Reggio Emilia, via G.Campi 213/A, I-41100 Modena, Italy.
[3] G.H.M.F.L. MPI-FKF and CNRS, BP 166, 38042 Grenoble, France. and
[4] Department of Chemistry, University of Manchester,
Oxford road, Manchester M13 9PL, United Kindom.
(Dated: 27 Sep 2004)

We investigate the role of topology and distortions in the quantum dynamics of magnetic molecules, using a cyclic spin system as reference. We consider three variants of antiferromagnetic molecular ring, i.e. $Cr_8$, $Cr_7Zn$ and $Cr_7Ni$, characterized by low lying states with different total spin $S$. We theoretically and experimentally study the low-temperature behavior of the magnetic torque as a function of the applied magnetic field. Near level crossings, this observable selectively probes quantum fluctuations of the total spin ("$S$ mixing") induced by lowering of the ideal ring symmetry. We show that while a typical distortion of a model molecular structure is very ineffective in opening new $S$-mixing channels, the spin topology is a major ingredient to control the degree of $S$ mixing. This conclusion is further substantiated by low-temperature heat capacity measurements.

Magnetic molecules are one of the best examples of ensemble of noninteracting quantum objects embedded in a solid state environment. Of particular relevance are molecules containing transition-metal ions whose spins are so strongly exchange-coupled that at low temperature each molecule behaves like a single-domain particle with fixed total spin $S$. Two major advantages in the research on these systems are on one side the outstanding degree of accuracy by which their magnetic dynamics can usually be modelled, and on the other side the opportunity to chemically engineer molecules possessing desired physical properties. Besides having obvious fundamental interest, the comprehension and control of the dynamics of these systems is necessary for implementing the envisaged technological applications as nanoscopic classical or quantum bits (qubits)[1–3].

Although the fixed-total-spin picture is sufficient to describe most low-temperature properties, there are subtle but remarkable effects characterizing the dynamics produced by *quantum* fluctuations of the total spin, the so-called "$S$-mixing" [4, 5]. Their consequences are not straightforward and are qualitatively different from those of thermal fluctuations. For instance, they hugely affect the tunneling of the magnetization in nanomagnets thus potentially preventing their use for classical information storage [5]. In the proposed quantum computing applications, $S$-mixing may wipe out the qubits through "leakage", i.e. dynamical evolution of the molecule out of the computational basis [3]. Therefore, a key goal is to understand how to tune $S$-mixing in the design of new molecules, and which aspects are the most critical and deserve care.

In this work, we show that the spin topology of the molecule is a major ingredient to control the degree of $S$-mixing. We consider three variants of an octanuclear antiferromagnetic (AF) ring, i.e. $Cr_8$ [6] and the chemically-substituted $Cr_7Zn$ and $Cr_7Ni$ [7]. While $Cr_8$ is topologically a ring, in the heterometallic compounds the ring topology is broken in two different ways, i.e. by magnetically opening the ring in $Cr_7Zn$ ($Zn^{2+}$ is nonmagnetic), and by changing the spin at the Ni site in $Cr_7Ni$. It must be stressed that $Cr_8$ is *not* an ideal ring, its point symmetry being lower than $C_8$. In particular, $Cr_8$ has only $C_4$ symmetry at room temperature, and moreover it displays hints of a symmetry lowering of the magnetic Hamiltonian al low $T$[6]. Recent X-ray experiments also evidence a structural transition at about 180 K, lowering the crystallographic ring symmetry to $C_2$[10].

In general, when the point group symmetry of a molecule is lowered, new $S$-mixing channels open up through the loosening of selection rules for intermultiplet matrix elements of the anisotropic terms in the Hamiltonian. Although the way selection rules change reflects the type of symmetry lowering and not the way such lowering occurs, the actual degree of breaching of these rules in real systems depends on the specific mechanism of symmetry breaking. The series of AF rings provides the unique opportunity to assess the impact on $S$-mixing of a departure from an optimal geometry by either a typical distortion or a topological change. By applying a magnetic field we can selectively target specific pairs of levels belonging to different $S$-multiplets which cannot mix in the ideal $C_8$ ring symmetry, but converge to the same irreducible representation whenever the ideal symmetry is lowered. We will show that in the distorted $Cr_8$ ring the departure from such symmetry does not produce any detectable mixing of these levels. Instead, the topological change in $Cr_7Zn$ and $Cr_7Ni$ opens this new $S$-mixing channel so effectively that in appropriate conditions even the behavior of some macroscopic observables is amaz-ingly changed.

The three molecules can be described by the following spin Hamiltonian:

$$H = \sum_i J_i \mathbf{s}_i \cdot \mathbf{s}_{i+1} + \sum_i d_i(s_z^2(i) - s_i(s_i+1)/3)$$
$$+ \sum_{i>j} \mathbf{s}_i \cdot \mathbf{D}_{ij} \cdot \mathbf{s}_j + \mu_B \sum_i g_i \mathbf{B} \cdot \mathbf{s}_i, \quad (1)$$

where $\mathbf{s}_i$ ($i=1,8$) are spin operators of the $i^{th}$ magnetic ion in the molecule ($s_i = 3/2$ for $Cr^{3+}$, $s_i = 0$ for non-magnetic $Zn^{2+}$, $s_i = 1$ for $Ni^{2+}$). The first term ($H_{Heis}$) is the isotropic Heisenberg exchange interaction. The second term describes the axial local crystal-fields (CFs), and the third term represents the dipolar anisotropic intra-cluster spin-spin interactions, with $z$ the axis perpendicular to the ring plane. The last term is the Zeeman coupling $H_{Zeeman}$ with an external field $\mathbf{B}$. All the parameters used in this work have been determined from the analysis of specific heat in magnetic field [3, 8, 13] and by the analysis of inelastic neutron scattering data [6, 9]. Since $H_{Heis}$ is invariant under rotations, it conserves the length $|\mathbf{S}|$ of the total spin $\mathbf{S} = \sum_i \mathbf{s}_i$. Instead, the anisotropic terms do not conserve this observable. Nevertheless, since $H_{Heis}$ is *largely dominant*, $|\mathbf{S}|$ is almost conserved, and the energy spectrum of $H$ is made of a series of level multiplets with an almost definite value of $|\mathbf{S}| = \sqrt{S(S+1)}$. By neglecting $S$-mixing, the ground manifold has $S=0$ in $Cr_8$, $S=3/2$ in $Cr_7Zn$, and $S=1/2$ in $Cr_7Ni$. The application of an external field produces a series of level crossings (LCs) at critical field values $B_c$'s, where the ground-state value of $S$ increases by unity (see Fig. 1).

In an ideal antiferromagnetic ring (i.e. $J_i = J$, $d_i = d$, $g_i = g$), for any value of $\mathbf{B}$ the two lowest eigenstates of $H_{Heis} + H_{Zeeman}$ (say $|0\rangle$ and $|1\rangle$) belong to different irreducible representations of the molecule point-group. Therefore, $\langle 0|H|1\rangle = 0$, and $|0\rangle$ and $|1\rangle$ are not mixed by the anisotropy and the magnetic field. This implies that when the applied field produces a LC between $|0\rangle$ and $|1\rangle$, this cannot be turned into an anticrossing (AC) by the anisotropy. This does not hold if a departure from the ideal symmetry occurs. Hence, the study of these LCs provides the opportunity to selectively address the problem of quantifying the increase of $S$-mixing due to the breaking of symmetry and topology.

The real $Cr_8$ ring has only $C_4$ symmetry, which lowers to $C_2$ below 180 K[10]. Nevertheless, as we will show, the mixing of $|0\rangle$ and $|1\rangle$ by the anisotropy remains negligible even at the level crossings. In the case of $Cr_7Zn$ and $Cr_7Ni$ the group is $C_2$ like in $Cr_8$, but the breaking of the ideal symmetry is much more profound. In fact, the ring topology itself is changed. This leads to a large mixing of $|0\rangle$ and $|1\rangle$, and to the appearance of a qualitatively different physics at the critical fields $B_c$'s. If the field direction lies outside the ring plane and is different from the $z$-axis, the mixing turns the LCs (involving levels belonging to different multiplets) into ACs (see Fig. 1). As the AC fields $\mathbf{B_c}$ are approached the multiplet mixing is enhanced. For $\mathbf{B} = \mathbf{B_c}$, the ground state wavefunction is approximately $(|0\rangle + |1\rangle)/\sqrt{2}$ and the total spin of each cluster displays quantum oscillations between states with different values of $S$, which therefore is no longer a good quantum number. It is worth to point out that the oscillation frequency $\nu_{AC}$ may be tuned from zero to much more than 20 GHz (corresponding approximately to an AC gap $\Delta_{AC}/k_B \sim 1$ K ) by changing the magnetic field orientation (see inset of Fig. 3). Thus, $\nu_{AC}$ may easily be made large enough to overcome the decoherence of the spin dynamics caused by interactions with phonons and nuclei.

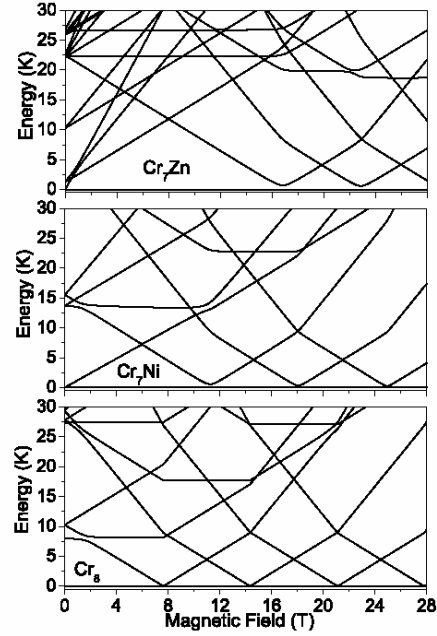

FIG. 1: Low-energy spectrum as function of the intensity of the applied field $B$. $\mathbf{B}$ forms an angle $\theta = 15^o$ with the $z$-axis. The ground state energy is set to zero for all values of $B$.

In order to prove experimentally the above picture, it is necessary to find an observable whose behavior enables these $S$-mixing effects to be identified unambiguously. It has been shown in[11, 12] that the magnetic torque ($\tau$) is such an observable. In fact, while in absence of $S$-mixing the torque signal as a function of $B$ is characterized by steps at the crossing fields, the oscillations of the total spin produce an additional sizeable peak-like contribution. Referring to a laboratory frame with $z'$ along the magnetic field direction, the unique $z$-axis perpendicular to the plane of the ring may rotate in the $z'x'$ plane forming an angle $\theta$ with $z'$ so the torque is exerted in the $y'$ direction. $\tau_{y'}$, hereafter simply $\tau$, can be written as proportional to $B\langle S_{x'}\rangle$. Quantum fluctuations of $|\mathbf{S}|$ near $B_c$ are accompanied by fluctuations of $S_{z'}$, and the latter

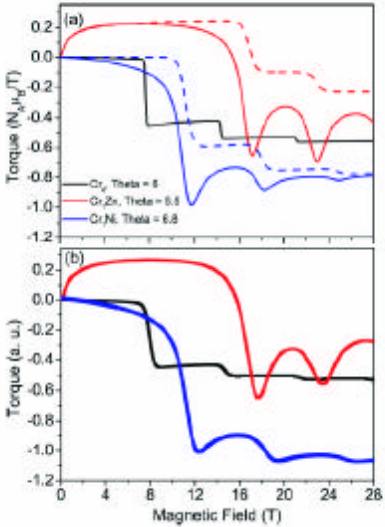

FIG. 2: (a): solid lines : calculated torque vs the applied field intensity $B$ for $Cr_8$ ($T = 50$ mK), $Cr_7Zn$ and $Cr_7Ni$ ($T = 400$ mK). Dashed lines : the same for $Cr_7Zn$ and $Cr_7Ni$ with $S$-mixing forced to zero. (b): Experimental results for the same conditions as in (a).

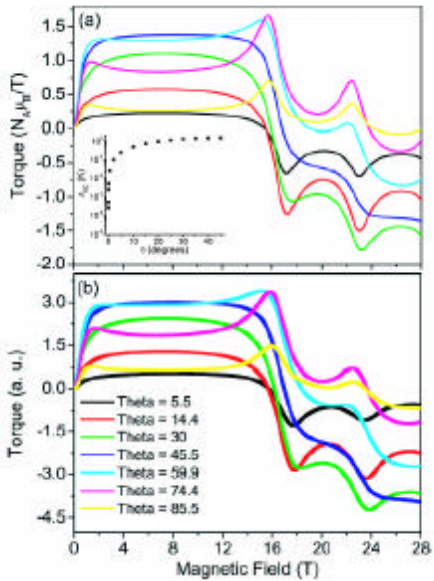

FIG. 3: (a): calculated torque vs the applied field intensity $B$ for $Cr_7Zn$ at $T = 400$ mK for different directions of the applied field. Inset : angular dependence of the first AC gap in $Cr_7Zn$. (b): Experimental results for the same conditions as in (a).

are connected with $\langle S_{x'} \rangle$. Indeed, near the ACs[11],

$$(\Delta S_{z'})^2 = \langle S_{z'}^2 \rangle - \langle S_{z'} \rangle^2 \simeq 0.25 \langle S_{x'} \rangle^2, \quad (2)$$

and therefore $\tau \propto 2B\Delta S_{z'}$. Accordingly, $\langle S_{x'} \rangle$ traces the increase and decrease of these fluctuations while sweeping over the AC, leading to a peak in the torque. In case of crossing $\Delta S_{z'}$ is always zero and the torque does not peak. This is evident in Fig. 2a, which reports the calculated $\tau(B)$ for **B** making an angle of $6°$ with the $z$-axis. While in the case of $Cr_8$ $\tau(B)$ is expected to display steps at the $B_c$'s, sharp peaks should occur in $Cr_7Zn$ and $Cr_7Ni$. By forcing $S$-mixing to zero in the calculation for $Cr_7Zn$ and $Cr_7Ni$ the peaks disappear and a step-like behavior like that of $Cr_8$ is recovered. Peaks in $\tau(B)$ are washed out by thermal fluctuations when $k_BT$ overcomes the AC gap.

In the following, we report experimental results obtained on the three systems. We used $\sim 0.2 \times 0.4 \times 0.4 mm^3$ sized single crystals with a flat surface that contains the plane of the rings. The experimental set up consists in a torque magnetometer made of a CuBe cantilever with a capacitance bridge detection. The magnetometer was mounted on rotator operating with a $^3$He or a top loading dilution refrigerator both inserted in the 28T magnet at the *High Magnetic Field Laboratory* in Grenoble. Single crystals, whose chemical formula is $[Cr_8F_8Piv_{16}]$ and $[\{Me_2NH_2\}\{Cr_7MF_8(O_2CCMe_3)_{16}\}]$, with $M=$ Ni, Zn respectively, were immersed in the cryogenic liquid to assure temperature stability during the field sweep. Sweep rate was between 200-400gauss/s and we just mention that no time dependent effects were observed by using these sweeping rates.

Fig.2b reports three measured torque curves for the three compounds. For an intuitive understanding of these curves, we briefly discuss results obtained for $Cr_8$. At low fields the torque signal is small but for B>8T and $\theta=6°$ the negative sign of the torque indicates that the magnetic field tends to align the plane of the crystal along the magnetic field direction, i.e. there is an easy plane of magnetization. Each step of the curve corresponds to a field-induced switch of the ground state, namely for $Cr_8$, from the $|S,M_S\rangle=|0,0\rangle$ to the $|1,-1\rangle$ state at $B_{c1}=7.9T$, from the $|1,-1\rangle$ to the $|2,-2\rangle$ state at $B_{c2}=14.7T$ and from the $|2,-2\rangle$ to the $|3,-3\rangle$ state at $B_{c3}=21.4T$. These results perfectly agree with those reported in Ref.[13], but the data plotted in Fig.2b were obtained at much lower temperature and higher field, i.e. 50mK and 28T, to reduce the thermal broadening (FWHM $\sim 0.3T$ at $B_{c1}$) and better show details of the step like behavior. In particular, these new results show well *the lack of peak* at the crossing fields thus greatly reducing the upper bound of hypothetical gaps $\Delta_{AC}$ in $Cr_8$. Notice that the lack of peaks behavior was also experimentally observed for $Fe_6$ and $Fe_{10}$ molecular ferric wheels [14], so it is clear that this depends neither on the number of the ring sites nor on the particular spin value ($Fe^{3+}$ has in fact $s_i=5/2$). Conversely, the predicted peaks in $\tau(B)$ for $Cr_7Zn$ and $Cr_7Ni$ were *clearly observed*, thus confirming that the breaking of the ring topology opens a new $S$-mixing channel between the ground and first excited states much more effectively than a distortion. As discussed in Ref.[13], the small slope of the $\tau(B)$ curve between two crossing, mostly evident for $Cr_7Ni$ but also exhibited by $Cr_8$, is also an effect of mixing of different $S$ states, i.e. $S$ and $S+2$. Yet, this mixing channel exists even in the ideal ring symmetry and therefore it provides no qualitative information on the effect of symmetry breaking. Indeed, in $Cr_8$ such effect is correctly reproduced by assuming the ideal $C_8$ symmetry (see Fig. 2).

In order to assess whether our model correctly captures the behavior of the wavefunctions and of the AC gaps as a function of the field direction $\theta$, we made a series of measurements of $\tau(B)$ for various values of $\theta$. In Fig. 3 we compare calculated and measured curves for $Cr_7Zn$. Notice the particular evolution of the peak as a function of the $\theta$ angle and focus, for instance, at $B_{c1} \sim 16T$. The peak is negative for $\theta < 45°$ and turns positive for $\theta > 45°$ slightly moving towards smaller fields. This peculiar $\theta$-dependence of both the position and the amplitude of the peaks is perfectly reproduced by our simulations. This is the case also for $Cr_7Ni$. We stress that this agreement is obtained with no adjustable parameters. In fact, as mentioned above, all the microscopic parameters of the spin Hamiltonian were independently fixed by specific heat and INS experiments.

In order to directly probe the AC gap and its closing when the field orientation is in the ring plane ($\theta = 90°$), we have performed heat capacity measurements on a $Cr_7Ni$ crystal (size $0.7 \times 0.7 \times 0.3 mm^3$) at fixed temperature by using the $ac$ technique [13]. Results are plotted in Fig.4 and compared with calculations. For $\theta = 25°$ the existence of the AC gap is clearly visible by a finite heat capacity at $B_c$. Conversely, for $\theta = 90°$ the Schottky anomaly related to the energy gap between the lowest lying levels vanishes at $B_c$, reflecting the closure of the AC gap. On the contrary, the heat capacity of $Cr_8$ vanishes at $B_c$ for any orientation of the field[13].

The comparison of the results for $Cr_8$ on the one hand, and for the two heterometallic rings on the other hand, reveals that the cluster topology is a critical factor in determining the amount of $S$-mixing in the spin wavefunctions, and which $S$-multiplets are actually mixed. In addition, the excellent agreement between theory and experiments shows the degree of precision by which the magnetic properties of these molecules can be modelled and controlled.

In conclusion, by studying three different variants of an AF ring ($Cr_8$,$Cr_7Zn$ and $Cr_7Ni$), we have shown the crucial role played by topology in determining whether symmetry-allowed $S$-mixing channels are actually open. While a distortion preserving the ring topology as in the case of $Cr_8$ does not lead to any detectable change of $S$-mixing, the breaking of the ring topology in the heterometallic rings unlocks new $S$-mixing channels so effectively that in appropriate conditions these change com-pletely the behavior of a macroscopic observable like the magnetic torque. Since $S$-mixing is detrimental to possible applications of magnetic molecules, an optimal topology should be a key ingredient in future engineering efforts.

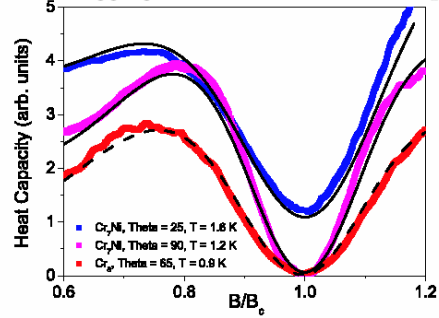

FIG. 4: Measured and calculated specific heat as function of $B/B_c$, with $B_c$ the first AC field, for $Cr_7Ni$ and two different field orientations. Data for $Cr_8$ from Ref.[13] have been added for comparison.

We would like to thank R. Caciuffo for helpful discussions. This work was partially financed by EC- RTN-QUEMOLNA contract nMRTN-CT-2003-504880. Measurements at GHMFL were supported by the EC "Access to research Infrastructure action of Improving Human Potential" Program.

$g_{Cr} = 1.98$ and $g_{Ni} = 2.2$[3].